\documentclass[pra,floatfix,twocolumn,showpacs,amsmath,amssymb]{revtex4}
 \usepackage{graphicx}
\usepackage{dcolumn}
\usepackage{bm}
\usepackage[usenames]{color}

\begin{document}

\title{Extracting the Mott gap from energy measurements in trapped atomic gases}
\author{Juan Carrasquilla$^{1}$ and Federico Becca$^{1,2}$}
\affiliation{
$^{1}$ International School for Advanced Studies (SISSA), Via Beirut 2, I-34151, Trieste, Italy \\
$^{2}$ Democritos Simulation Center CNR-IOM Istituto Officina dei Materiali, Trieste, Italy 
}
\date{\today}

\begin{abstract}
We show that the measure of the so-called {\it release energy}, which is an
experimentally accessible quantity, makes it possible to assess the value of
the Mott gap in the presence of the confinement potential that is unavoidable 
in the actual experimental setup. Indeed, the curve of the release energy as
a function of the total number of particles shows kinks that are directly
related to the existence of excitation gaps. Calculations are presented within
the Gutzwiller approach, but the final results go beyond this simple
approximation and represent a genuine feature of the real system. In the case
of harmonic confinement, the Mott gaps may be renormalized with respect 
to the uniform case. On the other hand, in the case of the recently proposed 
off-diagonal confinement, our results show good agreement with
the homogeneous case.
\end{abstract}


\maketitle

\section{Introduction}\label{sec:intro}

It has been established that ultra-cold Bose and Fermi gases trapped in optical
lattices provide experimental realizations of long-standing lattice models 
widely considered in condensed matter physics and statistical mechanics, such
as the Bose and Fermi Hubbard models~\cite{lewenstein,folling}. 
One of the most spectacular achievements was the observation of the superfluid
to Mott-insulator transition as the optical lattice depth is 
varied~\cite{bloch}. This transition has been explained in light of the 
Bose-Hubbard model with on-site repulsive interactions and hopping between 
nearest-neighboring sites. In the superfluid phase, each atom is spread out 
over the entire lattice, with long-range phase coherence, whereas in the 
insulating phase, exact numbers of atoms are localized at individual lattice 
sites, with no phase coherence across the lattice; this phase is characterized
by a gap in the excitation spectrum and a vanishing compressibility. 
Similarly, recent experiments have provided evidence on the formation of a 
Mott insulator of fermionic atoms in an optical lattice~\cite{jordens}. 
Several challenges arise when quantitatively comparing experimental data and 
theoretical results obtained from the analysis of infinite systems. One of the 
most important problems is due to inevitable spatial inhomogeneities induced 
by the optical trap, which is necessary to confine 
particles~\cite{batrouni,jaksch,wessel}. Further complications are connected 
to determine temperature effects present in experiments~\cite{pupillo}, and 
also to the limited available tools for the experimental characterization of 
the phases. 

Here, we would like to focus our attention on the possibility to make 
quantitative estimations of the Mott gap $E_g$ from relatively simple 
quantities that can be experimentally addressed. In presence of the optical 
trap, it is usually not possible to have a Mott insulating phase throughout 
the whole lattice and compressible regions intrude the system. This fact has 
been widely discussed both experimentally and theoretically and shows up 
through the typical ``wedding-cake'' profile of 
density~\cite{bloch,batrouni,jaksch,wessel,rigol,schroll}. 
Therefore, the system always possesses regions that are locally compressible 
and a precise determination of the Mott gap is subtler than in the 
homogeneous case. Some preliminary attempts to measure the excitation spectrum
of interacting bosons have been performed by using Bragg 
spectroscopy~\cite{stoferle}.
Other approaches to characterize the appearance of Mott-insulating regions from
experimentally accessible quantities have been also 
proposed~\cite{roscilde,rigolkink}. Here, we would like to propose an 
alternative approach that is based upon energy measurements only and could 
give important insights about the appearance of Mott-insulating regions, as 
well as the actual value of the gap. 

For an {\it infinite and homogeneous} system, the excitation gap $E_g$ can be 
calculated from the knowledge of the total energy for different particle 
numbers, namely $E_g = \mu^{+} - \mu^{-}$, where 
$\mu^{\pm} = \pm (E_{N \pm 1}-E_{N}$), $E_{N}$ being the ground-state energy 
with $N$ particles. The Mott gap $E_g$ is finite whenever $\mu^{+} \ne \mu^{-}$
and, therefore, introduces a discontinuity in the first derivative of the 
energy with respect to the density. 

In this paper, we will show that the existence of incompressible Mott regions 
and the values of the corresponding gaps can be obtained from so-called 
{\it release energy} $E^{rel}$, which may be  measured 
experimentally~\cite{dalfovo}. Indeed, $E^{rel}$ is obtained by integrating 
the momentum distribution of the atoms after having switched off the 
confinement and let the atoms expand freely. The release energy is given by 
the sum of the kinetic and interaction energies just before switching off the
trap~\cite{dalfovo}
\begin{equation}
E^{rel} = E^{kin} + E^{int}.
\end{equation}
We notice that it would be much more difficult to extract the 
{\it total energy}, $E^{tot} = E^{kin} + E^{int} + E^{pot}$ that also includes
the potential term due to the trap, since this would require the knowledge of 
the density profile in the presence of the trap, which is hard to reconstruct. 

Specifically, we address two types of confinement {\it i)} the usual harmonic 
confinement and {\it ii)} a recently proposed off-diagonal 
confinement~\cite{rousseau}. In such a confinement, the strength of the hopping
parameter is varied across the lattice, being maximum at the center of the 
lattice and vanishing at its edges, which naturally induces a trapping of the 
particles.

We consider bosons loaded in one- and two-dimensional optical lattices and 
use an insightful variational approach based upon the Gutzwiller wave 
function~\cite{rokhsar,krauth}. Similar results must hold also in fermionic 
systems. Moreover, although approximated, this variational method is expected
to correctly capture the behavior of the exact ground state. We find that the 
presence of Mott regions in the system is signaled by discontinuities in the 
derivative of the release-energy curve with respect to the total number of 
bosons, reminiscent of the presence of a gap in infinite homogeneous system.
In the case of harmonic confinement, the measured gap may be substantially
smaller than the one of the uniform system, whereas a much closer agreement
is achieved by considering the off-diagonal confinement.

The emergence of strong signatures due to the formation of Mott domains in 
the energy measurements has been explored in Ref.~\cite{rigolkink}, where it 
has been shown that the appearance of Mott domains is accompanied by minima 
in the release-energy curve as function of the on-site interaction parameter.
Here, we make a further step in this direction and provide a direct and
quantitative connection between energy calculations and Mott gaps, in 
presence of an external trap.

The paper is organized as follows: in section~\ref{sec:model}, we introduce
the Bose-Hubbard model and describe the variational method, in 
section~\ref{sec:results}, we present our results, and finally, in 
section~\ref{sec:conclusions}, we draw our conclusions.
 
\section{Model and method}\label{sec:model}

Our starting point is the Bose-Hubbard model which describes interacting 
bosons on a lattice~\cite{jaksch}:
\begin{equation}\label{hambose}
{\hat {\cal H}}=-\frac{1}{2} \sum_{\langle i,j \rangle} t_{i,j} 
{\hat b}^\dagger_i {\hat b}_j + h.c.
+ \frac{U}{2} \sum_i {\hat n}_i ({\hat n}_i-1) + \sum_i \epsilon_i {\hat n}_i,
\end{equation}
where $\langle \dots \rangle$ indicates nearest-neighbor sites, 
${\hat b}^\dagger_i$ (${\hat b}_i$) creates (destroys) a boson on site $i$, 
and ${\hat n}_i$ is the local density operator. $U$ is the on-site
interaction, $t_{i,j}$ is the hopping amplitude, and $\epsilon_i$ is 
a local energy offset due to an external trapping potential.

To study this model and describe its ground-state properties we use a 
mean-field approximation based on the Gutzwiller ansatz~\cite{rokhsar,krauth}.
This simple approach is able to capture important features of the true ground 
state and provides a qualitatively correct description of quantities such as 
the local density or the total energy, even in presence of spatial 
inhomogeneities~\cite{jaksch,rokhsar,krauth,schroll}. 
More involved wave functions with a long-range Jastrow factor~\cite{capello},
or numerically exact calculations~\cite{capogrosso} may be considered, but a 
much larger computational effort would be required. Instead, here we are just 
interested in showing that Mott gaps can be extracted from the behavior of
the release energy as a function of the total number of particles, and we
do not expect qualitative differences when considering more accurate 
approaches.

\begin{figure}
\includegraphics[width=\columnwidth]{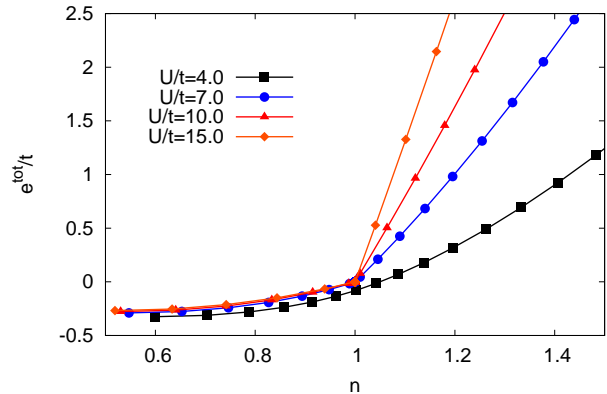}
\caption{\label{fig:clean}
(Color on-line) Energy per site versus density in the case of the 1D
homogeneous model for different values of $U/t$.}
\end{figure}

\begin{figure}
\includegraphics[width=\columnwidth]{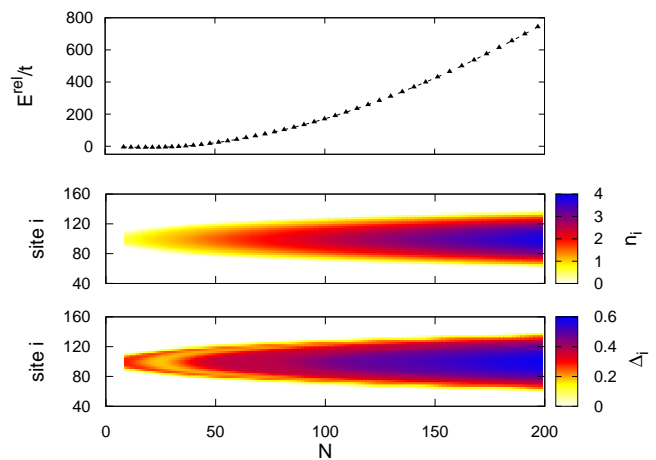}
\caption{\label{fig:U4}
(Color on-line) Upper panel: Release energy. Middle Panel: local density $n_i$
across the lattice sites. Lower Panel: local density fluctuations $\Delta_i$ 
across the lattice sites. A value of  $V_0/t=0.01$ is considered. 
All quantities are shown as a function of the total number of bosons $N$ and 
the interaction is set to $U/t=4$. Calculations are shown for the 1D model.}
\end{figure}

Within the Gutzwiller ansatz the ground-state wave function is approximated as
\begin{equation}\label{gutzwiller}
|\Psi_G \rangle = \prod_i \left ( 
\sum_{m=0}^{\infty} f_m^i |m \rangle_i \right ),
\end{equation}
where $|m\rangle_i$ is the Fock state with $m$ particles at site $i$ and 
$f_m^i$ are variational parameters which have to be determined by minimizing 
the expectation value of the Gutzwiller ansatz on the Hamiltonian in 
Eq.~(\ref{hambose}). The sum in Eq.~(\ref{gutzwiller}) runs from states with 
zero particles up to infinity; however, from a numerical point of view, we 
have to consider a cutoff and take only states up to a maximum number of 
particles (per site) $M_{max}^i \gg \langle \Psi_G|{\hat n_i}|\Psi_G\rangle$ 
such that the contribution of those states with higher density are negligible 
and observables are converged to a certain desired precision. 

Equivalently, the Gutzwiller wave function can be introduced as the ground 
state of the following mean-field Hamiltonian~\cite{sheshadri,buonsante}:
\begin{align}\label{meanfbose}
{\hat {\cal H}}_{mf}=-\frac{1}{2} \sum_{\langle i,j \rangle} t_{i,j}
\left( {\hat b}^\dagger_i \Psi_j + \Psi_i^* {\hat b}_j-\Psi_i^*\Psi_j \right) 
+ h.c. 
\nonumber \\
+ \frac{U}{2} \sum_i {\hat n}_i ({\hat n}_i-1) + 
\sum_i \left( \epsilon_i -\mu\right) {\hat n}_i,
\end{align}
where $\Psi_i$ is the mean-field potential which is self-consistently defined 
as $\Psi_i=\langle \Psi_G|{\hat b}_i|\Psi_G\rangle$; it can be shown that 
$f_m^i$ is related to the ground-state eigenvector components of the converged 
solution of the local Hamiltonian~(\ref{meanfbose})~\cite{sheshadri}.
The parameter $\mu$ is the chemical potential that fixes the number of bosons. 
In the recent past, the Gutzwiller approach has been successfully used 
to study a wide range of bosonic systems, like for instance disordered 
potentials~\cite{buonsante2}, boson-boson mixtures~\cite{buonsante3}, and even
time-dependent problems~\cite{zakrzewski,snoek,menotti}.

For the sake of simplicity and to simplify the presentation of the results, 
we first consider the one-dimensional (1D) case, which is the limiting case 
where a collection of non-interacting tubes is created. An almost 1D model 
may be easily generated by using different lasers in the three different 
spatial directions and has been experimentally 
considered~\cite{stoferle,fallani}. We mention that in 1D, within the 
Gutzwiller approach, the compressible phase has inevitably a finite condensate
fraction. However, this approach correctly reproduces the opening of the 
excitation gap at the Mott transition; therefore, the presence of a condensate 
does not imply relevant qualitative differences on the estimation of the gap 
with respect to an unbiased calculation. 
Furthermore, we also report some results for two dimensions (2D) for which we
obtain similar conclusions. We emphasize that all the results remain valid 
in higher spatial dimensions because the superfluid to Mott-insulator 
transition occurs in any dimension accompanied by the opening of a gap in the 
spectrum. 

We consider a lattice with a harmonic potential of the form 
$\epsilon_i=V_0 r_i^2$, where $r_i$ is the distance of site $i$ from the 
center of the lattice. In this case, the hopping amplitude is kept constant 
for all lattice sites $t_{i,j}=t$. In addition, we also analyze the case of 
off-diagonal confinement in which $\epsilon_i=0$ and 
$t_{i,i+1}=4t \times i \times (L-i)/L^2 $, where $L$ is the total number of
sites. In both cases, we evaluate local quantities which allow us to determine
whether a certain region across the lattice is in a compressible or 
incompressible state, as well as the release energy of the system. 
In particular, we will show, besides the release energy, the local density 
$n_i=\langle \Psi_G|{\hat n_i}|\Psi_G\rangle$ and its fluctuations
$\Delta_i= \langle \Psi_G|{\hat n}_i^2|\Psi_G\rangle - n_i^2$.

\begin{figure}
\includegraphics[width=\columnwidth]{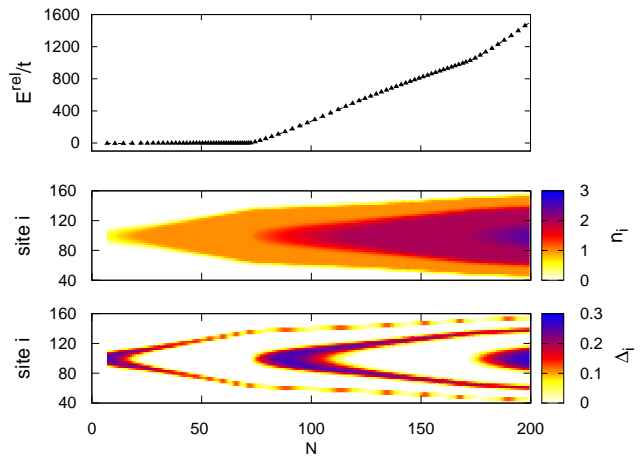}
\caption{\label{fig:U15}
(Color on-line) The same as in Fig.~\ref{fig:U4} for $U/t=15$.}
\end{figure}

\begin{figure}
\includegraphics[width=\columnwidth]{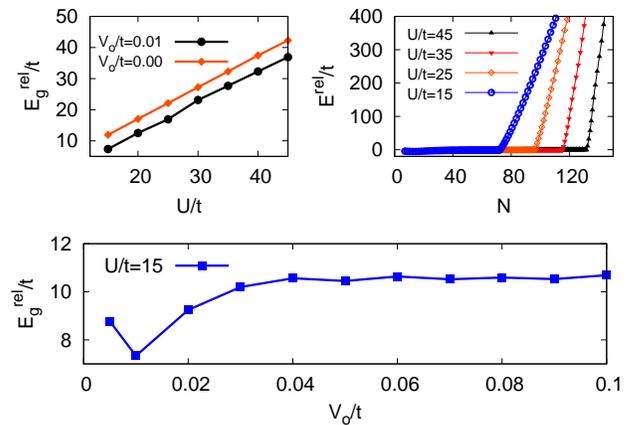}
\caption{\label{fig:summary}
(Color on-line) Upper panels: first Mott gap obtained from the release energy 
as function of $U/t$ for $V_0/t=0.01$, for comparison the homogeneous case
is also reported (left); release-energy curves for different $U/t$ and 
$V_0/t=0.01$. Lower Panel: first Mott gap as function of $V_0/t$ for $U/t=15$.
Calculations are shown for the 1D model.}
\end{figure}

\section{Results}\label{sec:results}

Before showing the results for the confined system, which is relevant for
experiments, let us briefly discuss the homogeneous case, with 
$\epsilon_i=0$ and $t_{i,j}=t$. In this case, a superfluid-Mott transition
takes place at integer fillings whenever the on-site interaction $U$ is large
enough. On the other hand, for any non-integer fillings the ground state is
always superfluid and, therefore, compressible. 
In Fig.~\ref{fig:clean}, we report the energy curve as a function of the 
density $n$. Within the Gutzwiller approximation the values of the critical 
interaction may be determined analytically, i.e., 
$U_c/t= D (\sqrt{n}+\sqrt{n+1})^2$, where $n$ is an integer~\cite{krauth}.
Whenever $U<U_c$, the energy curve is smooth with a positive curvature,
implying a finite compressibility and a vanishing gap. On the contrary, for
$U>U_c$, there is a discontinuity in the curve at integer fillings,
(the behavior in the vicinity of $n=1$ is reported in Fig.~\ref{fig:clean}),
signaling the presence of the Mott gap. The latter one can be estimated by 
considering the change of the slope close to the discontinuity.

\begin{figure}
\includegraphics[width=\columnwidth]{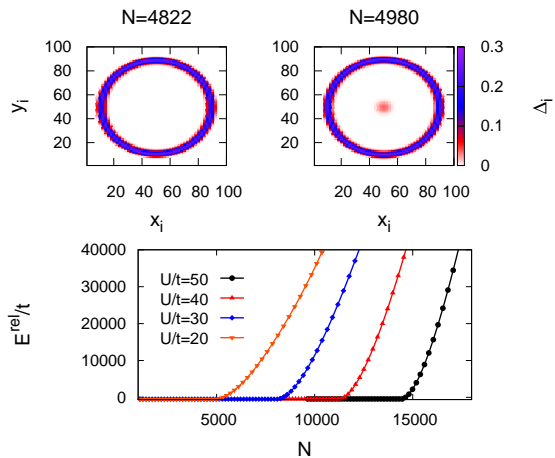}
\caption{\label{fig:2D}
(Color on-line) Upper panels: 2D results on local density fluctuations 
$\Delta_i$ across the lattice before (left) and after (right) the kink in the 
release energy for $U/t=20$. Lower panel: Release energy for 
different values of $U/t$ and fixed $V_0/t=0.01$.}
\end{figure}

Let us now switch on the harmonic confinement by setting $V_0/t=0.01$. 
In Figs.~\ref{fig:U4} and~\ref{fig:U15}, we present the results for the 
release energy as function of the total number of bosons $N$, as well as the 
local quantities $n_i$ and $\Delta_i$ across the lattice sites. For the case 
with $U/t=4$ (see Fig.~\ref{fig:U4}) there are no insulating phases, no matter
what the number of bosons is. The density profile is smooth, with a broad 
maximum at the center of the trap. In this case, all regions of the lattice are 
(locally) compressible and, therefore, the ground state is gapless.
This is not the case for $U/t=15$ (see Fig.~\ref{fig:U15}), where insulating 
regions are expected. Indeed, what is found is the usual ``wedding-cake'' 
structure in the local density: the Mott regions with integer $n_i$ and 
vanishing $\Delta_i$ are surrounded by compressible regions which are locally
gapless. We emphasize that the vanishing of $\Delta_i$ is consequence of the 
Gutzwiller approach; in a more accurate description of a Mott insulator, 
this quantity is indeed finite, though it is strongly reduced with respect 
to its value in the superfluid regions. Here, the derivative of the 
release-energy curve clearly exhibits discontinuities that are reminiscent of 
the presence of a Mott gap. This discontinuity takes place whenever a new 
compressible region appears at the center of the trap, on top of the underlying
Mott phase, see Fig.~\ref{fig:U15}. In this way, we can define an energy gap 
for the confined system exactly as in the homogeneous system, namely 
$E^{rel}_{g}=\mu_{rel}^{+}-\mu_{rel}^{-}$, where 
$\mu_{rel}^{\pm}=\pm(E^{rel}_{N \pm 1}-E^{rel}_{N})$ ($E^{rel}_{N}$ being the
release-energy with $N$ particles). We would like to stress that similar 
results may be obtained taking 
$\mu_{rel}^{\pm}=\pm(E^{rel}_{N \pm M}-E^{rel}_{N})$, with $M \ll N$, which
can be considered in experiments. For the case shown in 
Fig.~\ref{fig:U15} (i.e., for $V_0/t=0.01$ and $U/t=15$), we obtain that the
first Mott gap is $E^{rel}_{g}/t \simeq 7.2$, to be compared with the Mott
gap with $n=1$ of the homogeneous case that gives $E_{g}/t \simeq 11.8$. 
The reduction of the measured gap comes from the fact that the release energy
in the presence of the trap contains not only the information about the local 
creation of the new compressible region at the center of the trap, but also 
about all other sites of the lattice, which do not undergo the Mott transition.
Therefore, the effect is spatially averaged over regions that are locally 
compressible and incompressible. Nevertheless, the effect that originates 
from the central sites is visible and allows us to provide an estimate of the 
gap associated with such a transition.

A summary of the results is reported in Fig.~\ref{fig:summary}, where we show
$E^{rel}_g$ (for the first Mott gap) as a function of $U/t$ for $V_0/t=0.01$
and as a function of $V_0/t$ for $U/t=15$. We find that, when the harmonic
trap is increased, the Mott gap saturates to $E^{rel}_g/t \simeq 10.6$, which
is close to the value of the homogeneous system. The initial depletion of
the Mott gap as a function of $V_0$ is due to the presence of (large) regions 
of compressible sites close to the borders of the system; by further increasing 
$V_0$, these regions shrink and the gap eventually tends to approach the value
of the uniform case.

\begin{figure}
\includegraphics[width=\columnwidth]{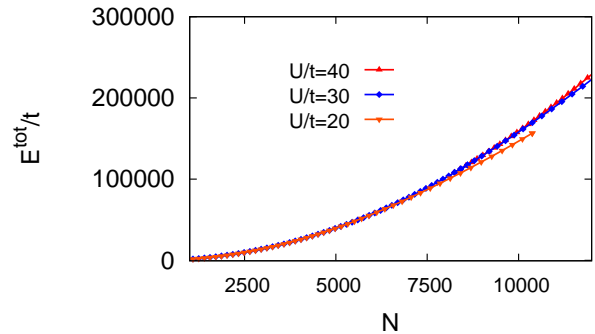}
\caption{\label{fig:totale}
(Color on-line) Results on total energy of the confined system in the 2D 
model as function of the number of particles for different values of $U/t$.}
\end{figure}

We now briefly consider the two-dimensional case, see Fig.~\ref{fig:2D}. 
Exactly as in 1D, in the appearance of a new compressible region at the center
of the trap is accompanied by a discontinuity in the derivative of the 
release-energy curve. Similar conclusions to those in 1D are obtained, 
confirming the appearance of kinks in the release-energy curve in 2D where 
the Gutzwiller mean-field approach is more reliable.

We stress that in presence of a diagonal confinement is such that compressible 
and incompressible regions coexist. Indeed, the true gap of the overall 
confined system will be zero, as the global compressibility is always 
finite~\cite{batrouni,roscilde}. Indeed, the total energy, as opposed to the 
release energy, is a completely smooth curve with positive curvature
(hence a vanishing real gap and finite compressibility). The total-energy 
curve as function of the total number of particles is shown in 
Fig.~\ref{fig:totale} for a 2D system. This curve is completely smooth, even 
when Mott insulating domains are present in the trap~\cite{batrouni}.

\begin{figure}
\includegraphics[width=\columnwidth]{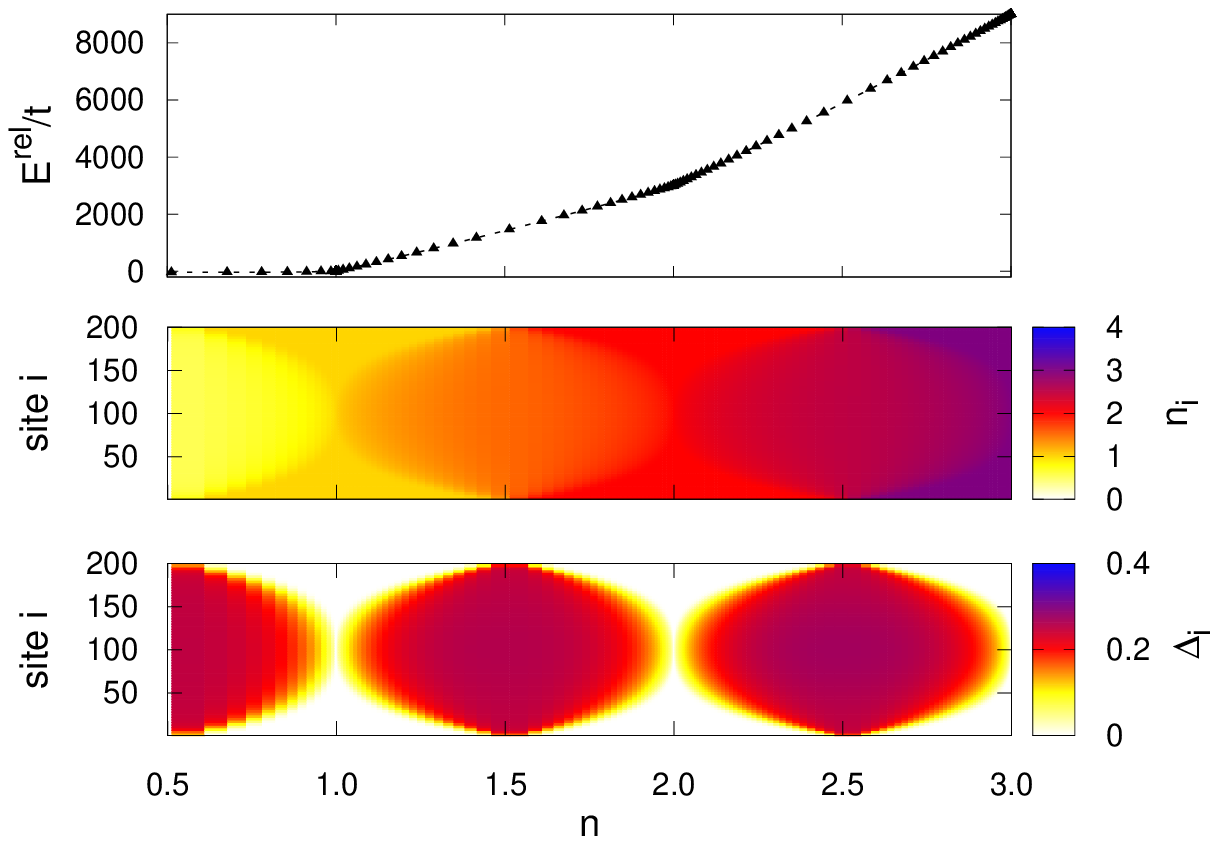}
\caption{\label{fig:ODCU15}
(Color on-line) The same as in Fig.~\ref{fig:U15} but for an off-diagonal
confinement. The number of sites is $L=200$ and $n=N/L$ is the density.}
\end{figure}

Now, let us analyze the case of off-diagonal confinement, recently proposed by
Rousseau and collaborators~\cite{rousseau}. In the experimental setup, the 
release-energy measurement corresponds to the actual total energy of the 
trapped system, since in this case there is no potential energy. 
In Fig.~\ref{fig:ODCU15}, we present the results for $U/t=15$. We observe that
the kinks in the energy curve occur exactly at values of the total number of 
bosons $N$ which are commensurate with $L$, i.e., for integer densities 
$n=N/L$, as in a homogeneous case. The advantage of this kind
of confinement is that it is possible to have a Mott-insulating phase 
throughout the whole lattice, as discussed in Ref.~\cite{rousseau}. 
By using the energy curve, we find that the value of the gap at $n=1$ is 
$E^{rel}_{g}/t \simeq 11.9$, remarkably close to the corresponding value in 
the homogeneous system ($E_g/t \simeq 11.8$) but slightly larger. 
This is  due to the fact that regions towards the border of the lattice 
are effectively in a deeper Mott phase due to their reduced kinetic energy. 
Away from integer densities, the system shows coexistence of compressible 
and incompressible states as in the harmonic confinement, with compressible 
sites always at the center of the trap and Mott sites on the borders, 
again due to the reduced kinetic energy there. 
 
\section{Conclusions}\label{sec:conclusions}

We have shown that relatively simple energy measurements may give important 
insights into the actual value of the Mott gap. In particular, it has been 
shown that the presence of the harmonic potential may renormalize the
value of the gap with respect to the uniform case. On the other hand, much
closer results may be achieved when the off-diagonal confinement will be 
experimentally realized. Although the experimental determination of the Mott
gap from release-energy curves may be difficult, since small error bar on both
energies and particle numbers are required, we hope that in the near future
this method will be successfully applied.

We acknowledge useful discussions with M. Fabrizio, L. Fallani, C. Fort, 
S. Giorgini, V. Rousseau, and A. Trombettoni.


\begin{thebibliography}{99}
\bibitem{lewenstein} M. Lewenstein, A. Sanpera, V. Ahufinger, B. Damski, 
   A. Sen De, and U. Sen, Adv. Phys. {\bf 56}, 243 (2007).
\bibitem{folling} M. Greiner and S. Folling, 
   Nature (London) {\bf 453}, 736 (2008).
\bibitem{bloch} M. Greiner, O. Mandel, T. Esslinger, T.W. Hansch, I. Bloch, 
   Nature (London) {\bf 415}, 39 (2002).
\bibitem{jordens} R. Jordens, N. Strohmaier, K. Gunter, H. Moritz, and
   T. Esslinger,  Nature (London) {\bf 455}, 204 (2008).
\bibitem{batrouni} G.G. Batrouni, V. Rousseau, R.T. Scalettar, M. Rigol,
   A. Muramatsu, P.J.H. Denteneer, and M. Troyer, 
   \prl {\bf 89}, 117203 (2002).
\bibitem{jaksch} D. Jaksch, C. Bruder, J.I. Cirac, C.W. Gardiner, 
   and P. Zoller,
   \prl {\bf 81}, 3108 (1998).
\bibitem{wessel} S. Wessel, F. Alet, M. Troyer, and G.G Batrouni, 
   \pra {\bf 70}, 053615 (2004).
\bibitem{pupillo} G. Pupillo, C.J. Williams, and N.V. Prokof'ev, 
   \pra {\bf 73}, 013408 (2006).
\bibitem{rigol} M. Rigol, G.G. Batrouni, V.G. Rousseau, and R.T. Scalettar,
   \pra {\bf 79}, 053605 (2009).
\bibitem{schroll} C. Schroll, F. Marquardt, and C. Bruder, 
   \pra {\bf 70}, 053609 (2004).
\bibitem{stoferle} T. St\"oferle, H. Moritz, C. Schori, M. Kohl, and 
   T. Esslinger, 
   \prl {\bf 92}, 130403 (2004).
\bibitem{roscilde} T. Roscilde, New J. Phys. {\bf 11}, 023019 (2009).
\bibitem{rigolkink} M. Rigol, R. T. Scalettar, P. Sengupta, and G. G. Batrouni,
   \prb  {\bf 73}, 121103(R) (2006)
\bibitem{dalfovo} F. Dalfovo, S. Giorgini, L.P. Pitaevskii, and S. Stringari,
   \rmp {\bf 71}, 463 (1999).
\bibitem{rousseau} V.G. Rousseau, G.G. Batrouni, D.E. Sheehy, J. Moreno, 
   and M. Jarrell, 
   \prl {\bf 104}, 167201 (2010).
\bibitem{rokhsar} D.S. Rokhsar and B.G. Kotliar, 
   \prb {\bf 44}, 10328 (1991).
\bibitem{krauth} W. Krauth, M. Caffarel, and J. Bouchaud, 
   \prb {\bf 45}, 3137 (1992).
\bibitem{capello} M. Capello, F. Becca, M. Fabrizio, and S. Sorella,
   \prl {\bf 99}, 056402 (2007); \prb {\bf 77}, 144517 (2008).
\bibitem{capogrosso} B. Capogrosso-Sansone, N.V. Prokof'ev, and B.V. Svistunov,
   \prb {\bf 75}, 134302 (2007); B. Capogrosso-Sansone, S. Gunes Soyler,
   N. Prokof'ev, and B. Svistunov, \pra {\bf 77}, 015602 (2008).
\bibitem{sheshadri} K. Sheshadri, H.R. Krishnamurthy, R. Pandit, and 
   T.V. Ramakrishnan, 
   Europhys. Lett. {\bf 22}, 257 (1993).
\bibitem{buonsante} P. Buonsante, F. Massel, V. Penna, and A. Vezzani, 
   \pra {\bf 79}, 013623 (2009). 
\bibitem{buonsante2} P. Buonsante, V. Penna, A. Vezzani, and P.B. Blakie,
   \pra {\bf 76}, 011602(R) (2007).
\bibitem{buonsante3} P. Buonsante, S.M. Gianpaolo, F. Illuminati, V. Penna, 
   and A. Vezzani, \prl {\bf 100}, 240402 (2008).
\bibitem{zakrzewski} J. Zakrzewski, \pra {\bf 71}, 043601 (2005).
\bibitem{snoek} M. Snoek and W. Hofstetter, \pra {\bf 76}, 051603(R) (2007).
\bibitem{menotti} C. Trefzger, C. Menotti, and M. Lewenstein, \pra {\bf 78},
   043604 (2008).
\bibitem{fallani} V. Guarrera, N. Fabbri, L. Fallani, C. Fort, 
   K.M.R. van der Stam, and M. Inguscio, 
   \prl {\bf 100}, 250403 (2008). 
\end{thebibliography}
\end{document}